\documentclass[10pt,letterpaper,twocolumn]{article} 

\usepackage{times,mathptmx}
\usepackage{ol2}
\usepackage{amsmath,amssymb}
\usepackage{color}

\begin{document}

\twocolumn[ 

\title{Observation of dispersive wave emission by temporal cavity solitons}

\vskip-3mm

\author{Jae K. Jang,$^{*}$ Miro Erkintalo, Stuart G. Murdoch, and St\'{e}phane Coen}

\affiliation{Physics Department, The University of Auckland, Private Bag 92019, Auckland 1142, New Zealand\\
             $^*$Corresponding author: jake.jang.ur.mate@gmail.com}

\begin{abstract}%
  We examine a coherently-driven, dispersion-managed, passive Kerr fiber ring resonator and report the first direct
  experimental observation of dispersive wave emission by temporal cavity solitons. Our observations are in
  excellent agreement with analytical predictions and they are fully corroborated by numerical simulations. These
  results lead to a better understanding of the behavior of temporal cavity solitons under conditions where
  higher-order dispersion plays a significant role. Significantly, since temporal cavity solitons manifest
  themselves in monolithic microresonators, our results are likely to explain the origins of spectral features
  observed in broadband Kerr frequency combs.
\end{abstract}

\noindent\ocis{(190.5530) Pulse propagation and temporal solitons; (190.4370) Nonlinear optics, fibers; (140.4780)
Optical resonators.}


] 

\noindent The emission of dispersive waves (DWs) by intense light pulses perturbed by high-order dispersion has been
studied for decades, and it is well documented in nonlinear fiber optics~\cite{wai_nonlinear_1986,
wai_radiations_1990, akhmediev_cherenkov_1995}. It typically occurs when the phase of a propagating pulse evolves
identically to the phase of a linear wave (or waves) at a different frequency, allowing this wave to be amplified
through resonant energy transfer. The vast majority of studies associate DW emission exclusively to bright solitons,
but the process is in fact much more general: it can occur with dark solitons \cite{karpman_stationary_1993} and even
with non-solitonic pulses both in the normal and in the anomalous dispersion regimes \cite{erkintalo_cascaded_2012,
webb_generalized_2013, conforti_dispersive_2013}. Today the process is recognized as one of the central frequency
conversion mechanisms in nonlinear fiber optics, where it has been harnessed to realize tunable sources of coherent
radiation~\cite{chang_highly_2010, mak_tunable_2013}, and identified as a critical component in the generation of
broadband supercontinua \cite{dudley_supercontinuum_2006, skryabin_colloquium:_2010}.

DWs have now also started to be put forward in another broadband frequency generation paradigm, namely the generation
of frequency combs from high-Q Kerr microresonators. This topic has been under intense experimental investigation
since 2007~\cite{delhaye_optical_2007, kippenberg_microresonator-based_2011}, and recent progresses in theoretical
modeling have identified some spectral features of these so-called ``Kerr combs'' as possibly arising from DW
emission \cite{erkintalo_cascaded_2012, coen_modeling_2013, lamont_route_2013}. More specifically, in some
conditions, Kerr combs are now being understood to be underlined in the time domain by dissipative structures known
as temporal cavity solitons (CSs)~\cite{leo_temporal_2010, coen_modeling_2013, coen_universal_2013}. It is these
short CS pulses, and the related periodic modulational instability (MI) patterns, that appear to emit DWs under low
dispersion conditions, with temporal instabilities --- when present --- responsible for additional spectral
broadening of the DW peak \cite{okawachi_octave-spanning_2011, erkintalo_coherence_2014}. This proposition is further
supported by the fact that DW emission can be linked to cascaded four-wave mixing~\cite{erkintalo_cascaded_2012}, the
frequency-domain process underlying Kerr comb generation~\cite{delhaye_optical_2007,
kippenberg_microresonator-based_2011}.

Temporal CSs have just been reported in Kerr comb microresonator experiments \cite{herr_temporal_2014} but an
inconvenient fact remains: temporal CSs have never been observed to emit DWs in any platform. This shortfall can be
explained by noting that, so far, all experiments conclusively involving temporal CSs have been carried out far from
any zero-dispersion wavelength (ZDW)~\cite{leo_temporal_2010, jang_ultraweak_2013, herr_temporal_2014}, i.e., in
conditions where higher-order dispersion plays a negligible role \cite{dudley_supercontinuum_2006,
skryabin_colloquium:_2010}. In this Letter, we address this point by studying experimentally temporal CSs, for the
first time in a low-dispersion cavity. Our study is performed in a single-mode fiber loop, in which we can more
reliably control the excitation of temporal CSs and the dispersion characteristics of the resonator compared to a
microresonator. DWs are positively identified in our experiment and their emission wavelength found to agree with a
recent theoretical model \cite{milian_soliton_2014}.

\begin{figure}[b]
  \centerline{\includegraphics[width=8.5cm]{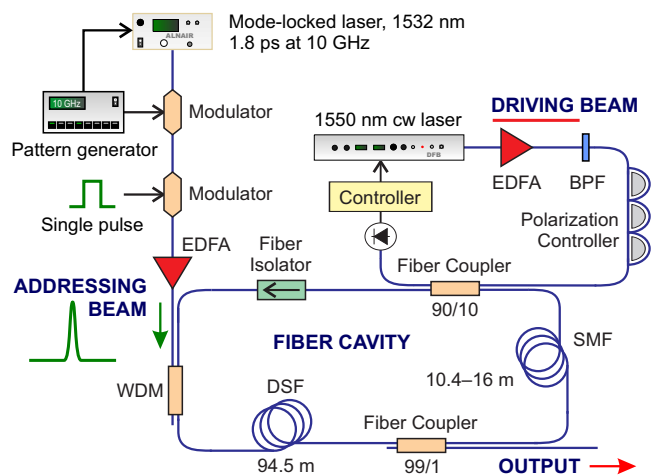}}
  \caption{A schematic of our experimental set-up. EDFA: Erbium-doped fiber amplifier, BPF: band-pass filter,
    WDM: wavelength division multiplexer, SMF: single-mode fiber, and DSF: dispersion-shifted fiber.}
  \label{setup}
\end{figure}
Our experimental set-up is illustrated in Fig.~\ref{setup} and is similar to that used in previous temporal CS
experiments \cite{leo_temporal_2010, jang_ultraweak_2013}. The cavity is coherently driven by a continuous-wave (cw)
ultra-narrow linewidth ($< 1$~kHz) DFB fiber laser at 1550~nm wavelength, which is amplified to 1--$1.3$~W. A
band-pass filter (BPF) rejects the noise from the optical amplifier before the driving beam is injected into the
cavity through a 90/10 fiber coupler. Our cavity incorporates an optical fiber isolator which prevents the build-up
of stimulated Brillouin scattering (SBS), a wavelength division multiplexer (WDM) to couple the addressing pulses
used to excite the CSs, a 99/1 tap coupler through which the output spectra are monitored with an optical spectrum
analyzer, and some extra length of fiber. The overall cavity finesse $\mathcal{F}=\pi/\alpha$ is about 18, with
$\alpha = 0.176$ representing half the percentage of power loss per round-trip. A PID feedback control loop actuating
the laser frequency locks the driving laser near a cavity resonance and keeps the system interferometrically stable.
The error signal of the control loop is derived from the part of the driving power that is reflected off the input
coupler. In practice we operate the laser with a detuning $\delta_0=0.5$--$0.56$~rad before the closest cavity
resonance. To excite temporal CSs, we used the same incoherent writing process as used in previous experiments
\cite{leo_temporal_2010, jang_ultraweak_2013}, namely cross-phase modulation between the cw intracavity field and
external addressing pulses. The addressing pulses were $1.8$~ps long and generated by a 1532~nm-wavelength
mode-locked laser with a 10~GHz repetition rate. These pulses were selected by a series of two intensity modulators
(see \cite{jang_ultraweak_2013}) before being amplified and launched in the cavity through the WDM coupler. For each
measurement, the cavity was filled with about 1300 temporal CSs. As all the CSs that circulate in our cavity are
identical \cite{leo_temporal_2010}, this simply boosts the spectral power of the CS component relative to the cw
background, improving the quality of our measurements.

The fiber path around the cavity is composed of $94.5$~m of dispersion-shifted fiber (DSF) and 10.4~m to 16.0~m of
standard single-mode fiber (SMF). Adjusting the length of the SMF segment allows us to vary the \textit{average}
cavity dispersion, and to explore CS dynamics in the regime where the third-order dispersive perturbation plays a
significant role. The combination of two fibers is justified by the relatively high-finesse: it implies that the
field does not vary significantly over the course of a single round-trip (the cavity photon lifetime being much
larger than the round-trip time) and therefore that the lumped aspect of the cavity is not so relevant. The
round-trip-averaged dispersion and nonlinearity are instead the important quantities \cite{leo_nonlinear_2013}. At
our operating wavelength, the DSF exhibits normal group-velocity dispersion (GVD), with 2nd- and 3rd-order dispersion
coefficients $\beta_2^\mathrm{DSF} = 1.95\ \mathrm{ps^2/km}$ and $\beta_3^\mathrm{DSF} = 0.2\ \mathrm{ps^3/km}$,
while the SMF presents anomalous dispersion with the well-known values $\beta_2^\mathrm{SMF} = -20\ \mathrm{ps^2/km}$
and $\beta_3^\mathrm{SMF} = 0.1\ \mathrm{ps^3/km}$. Based on these parameters, varying the length of the SMF segment
changes the average GVD of our cavity $\langle\beta_2\rangle$ from $-0.23$ to $-1.23\ \mathrm{ps^2/km}$. At the same
time $\langle\beta_3\rangle$ stays relatively constant (to better than 3\,\%) at $0.19\ \mathrm{ps^3/km}$. Clearly,
the \textit{relative} contribution of third-order dispersion with respect to second-order dispersion can thus be
controlled in our system. To quantify this, we define a normalized third-order dispersion coefficient
$d_3=\sqrt{2\alpha/L}\,\beta_3/(3|\beta_2|^{3/2})$ where $L$ is the total length of the cavity. The normalization is
such that the cavity field decay rate and the GVD coefficient are both set to $-1$
\cite{parra-rivas_third-order_2014, xu_experimental_2014}. When $d_3\ll 1$ (respectively, $\gg1$), 2nd (3rd) order
dispersion dominates, and when $d_3\sim 1$ the two contribute comparably. Finally, we estimate the average nonlinear
coefficient of the cavity to be about $\gamma = 1.9\ \mathrm{W^{-1}\ km^{-1}}$.

In addition to experiments we also numerically simulate the nonlinear cavity dynamics. Our simulations use a full
lumped model, in which each cavity element is separately taken into account. Propagation through each fiber segment
is modeled using a generalized nonlinear Schr\"odinger equation~\cite{dudley_supercontinuum_2006}, the driving field
is interferometrically added at the beginning of each roundtrip, and the remaining discrete losses and/or components
are modeled through the use of localized transmission functions.

\begin{figure}[t]
  \centerline{\includegraphics{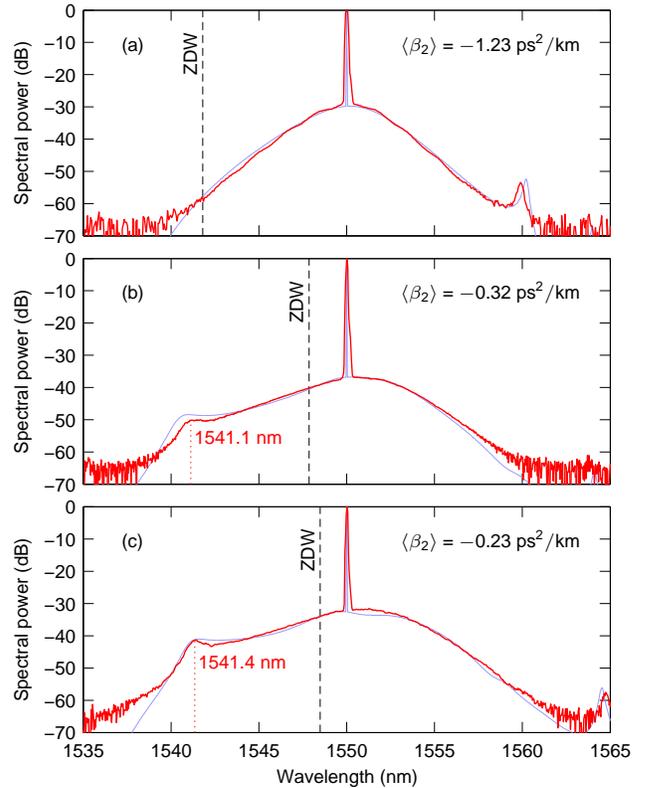}}
  \caption{Experimental (red) and numerical (blue) spectra of temporal CSs for three different values of average
    cavity dispersion as shown in each figure. In each case, the black dashed line indicates the position of the ZDW
    while the red dotted line highlights the observed DW wavelength. (a) $P_\mathrm{in}=1$~W, $\delta_0=0.55$~rad;
    (b) $P_\mathrm{in}=1$~W, $\delta_0=0.5$~rad; (c) $P_\mathrm{in}=1.3$~W, $\delta_0=0.56$~rad.}
  \label{fig:spectra}
  \vskip-3mm
\end{figure}
Figures~\ref{fig:spectra}(a)--(c) show experimental (red) and numerical (blue) spectra of temporal CSs obtained for
three different averaged cavity dispersion. We first note the overall remarkable agreement between the simulations
and the experiments. Results in Fig.~\ref{fig:spectra}(a) correspond to a cavity with 16~m of SMF, or equivalently an
average 2nd-order dispersion coefficient $\langle\beta_2\rangle = -1.23\ \mathrm{ps^2/km}$. The spectrum is typical
of a temporal CS: the broad spectral wings are associated with the narrow temporal pulsed structure of the soliton,
while the central narrow ``DC'' peak highlights the presence of the cw background on which temporal CSs are
superimposed \cite{leo_temporal_2010}. Although the dispersion is already relatively low, the CS spectrum does not
extend significantly past the ZDW (highlighted as a dashed line in Fig.~\ref{fig:spectra}), and no significant
spectral features are present in the normal dispersion regime where DWs would be expected. This is attributed to the
fact that the contribution from 3rd-order dispersion is still small compared to $\beta_2$, and indeed the normalized
3rd-order dispersion coefficient is very small, $d_3 \simeq 0.08 \ll 1$. The only noticeable feature seen in
Fig.~\ref{fig:spectra}(a) is a quasi-phase-matched Kelly sideband at 1560~nm in the anomalous dispersion regime. It
arises due to periodic nature of our cavity which is made up of two different fiber types
\cite{kelly_characteristic_1992}. Consequently, we note that the mean-field model that is usually employed to
describe CSs \cite{lugiato_spatial_1987, leo_temporal_2010, coen_modeling_2013} does not account for this Kelly
sideband. A full lumped cavity model as we have been using is required instead.

As the average dispersion is reduced to $-0.32\ \mathrm{ps^2/km}$ (corresponding to $10.9$ m of SMF, and to $d_3
\simeq 0.64$), the spectrum develops a peak at $1541.1$~nm in the normal dispersion regime
[Fig.~\ref{fig:spectra}(b)]. The peak shifts slightly to $1541.4$~nm and becomes more prominent
[Fig.~\ref{fig:spectra}(c)] at an even lower average dispersion of $-0.23\ \mathrm{ps^2/km}$ (corresponding to
$10.4$~m of SMF, and to $d_3 \simeq 1.05$). This peak is a clear signature of DWs emitted by the circulating temporal
CSs, and the trend observed when the average dispersion reduces fully agrees with this conclusion. Indeed, as we
reduce the average GVD coefficient $\langle\beta_2\rangle$, we are effectively shifting the ZDW towards the driving
beam wavelength. This pushes the DW towards the driving beam, leading to a more efficient amplification
\cite{akhmediev_cherenkov_1995}.

To more quantitatively establish that the spectral feature seen in Figs.~\ref{fig:spectra}(b)--(c) corresponds to a
DW, we must show that its wavelength obeys the relevant phase-matching condition. In some cases, especially when
broadband Kerr comb generation takes place in high-Q microresonators, the known condition for conservative solitons
\cite{akhmediev_cherenkov_1995} has been shown to provide a good approximation \cite{coen_modeling_2013}. However,
this limit does not apply here and a more general prediction requires the cavity geometry, and in particular the
effect of the cavity detuning, to be fully taken into account. The theoretical study of
Ref.~\cite{milian_soliton_2014} has recently addressed this issue based on the following argument: In the cavity
configuration, because of the dissipative losses, the DW is a stationary localized radiation tail attached to the CS.
The DW frequency can thus be found by examining how this tail asymptotically approaches the cw background on which
the temporal CSs are superimposed. Specifically, one considers the ansatz $E=E_0 + a \exp(-i Q \tau) + b \exp(i Q^*
\tau)$ (the $*$ denoting complex conjugation) in the mean-field model of the cavity \cite{lugiato_spatial_1987,
leo_temporal_2010, coen_modeling_2013, chembo_spatiotemporal_2013}, with $\tau$ the fast-time describing the temporal
profile of the field inside the cavity and $E_0$ the cw lower state background, and one looks for a self-consistent,
stationary, linearized ($|a|,|b| \ll |E_0|$) DW solution. This leads to the following condition for the complex
frequency $Q$:
\begin{multline}
  -\alpha + i \frac{\beta_3 L}{3!} Q^3 - i V Q \\
  \pm i \sqrt{\left(2\gamma L P_0 - \delta_0 + \frac{\beta_2 L}{2!} Q^2 \right)^2-(\gamma L P_0)^2} = 0\,.
  \label{eq:dw}
\end{multline}
Here $P_0=|E_0|^2$ is the power level of the cw lower state background, while $V$ is a drift velocity that accounts
for the fact that 3rd-order dispersion (or any odd order of dispersion) makes the CSs group-velocity slightly
different from that of the driving field \cite{leo_nonlinear_2013, parra-rivas_third-order_2014}. This is associated
with a spectral shift of the CS spectrum with respect to the driving frequency and in \cite{milian_soliton_2014} has
been interpreted as resulting from the spectral recoil due to the emission of DWs. Specifically, $V$ represents the
group-delay accumulated by the temporal CSs with respect to the driving field over one round-trip (and has the units
of time). We note that the above equation can be generalized to include arbitrary orders of dispersion by adding the
extra even (odd) order dispersion terms in the form $\beta_k L Q^k/k!$ (for $k\ge 4$) to the $\beta_2$ ($\beta_3$)
contributions, respectively.

In considering Eq.~(\ref{eq:dw}), we can restrict ourselves to the $+$ sign alternative in front of the square root
because the solutions of the two cases are related by the transformation $Q \rightarrow -Q^*$. This fact highlights
that, in the cavity geometry, DW peaks should appear in pairs symmetrically located with respect to the driving
frequency. Further analysis of the eigenvectors of the problem \cite{milian_soliton_2014} indicates however that in
our experiment the dominant DW is always up-frequency shifted. The down-frequency shifted symmetrical counterpart is
predicted to be significantly weaker, and so far we have been unable to detect it.

Equation~(\ref{eq:dw}) was studied in \cite{milian_soliton_2014} with the loss coefficient $\alpha$ neglected, which
was appropriate for the context of high-Q microresonators pertinent to that work. In that condition, the
solutions~$Q$ are purely real (and directly interpreted as the angular frequency shift of the DW with respect to the
driving frequency). In our fiber cavity experiment, the losses are more important but we can still simplify
Eq.~(\ref{eq:dw}) by noting that with our measured background power level $P_0 \simeq 500$--$600$~mW, the nonlinear
round-trip phase shift $\gamma L P_0 \sim 0.1$~rad, making the second-term in the square root, $(\gamma L P_0)^2$,
negligible. Eq.~(\ref{eq:dw}) then becomes a polynomial in $Q$,
\begin{equation}
  \frac{\beta_3 L}{3!} Q^3 + \frac{\beta_2 L}{2!} Q^2 - V Q + \left[ (2 \gamma L P_0-\delta_0) + i \alpha \right] = 0\,.
  \label{eq:dw_simplified}
\end{equation}
In the form above, Eq.~(\ref{eq:dw_simplified}) can be easily compared to the classical DW phase-matching condition
of conservative solitons \cite{wai_radiations_1990, akhmediev_cherenkov_1995} and the differences with the cavity
geometry readily identified. We note in particular that the peak power of the temporal CS does not play any direct
role: rather, it is the power of the CS background $P_0$ which is relevant. When the dispersion terms $\beta_k$
dominate, Eq.~(\ref{eq:dw_simplified}) approaches the conservative soliton expression \cite{coen_modeling_2013}.

In order to solve Eq.~(\ref{eq:dw_simplified}) for the parameters of our experiment, the only thing we lack is the CS
drift velocity~$V$. We can obtain this quantity from the observed spectral recoil of the CS spectrum. In
Fig.~\ref{fig:spectra}(c), the peak of the CS spectral components does not coincide with the driving wavelength but
rather is shifted towards longer wavelengths by about 1~nm [corresponding to a frequency shift
$\Delta\Omega_\mathrm{CS}/(2\pi) = -125$~GHz]. This shift translates into an extra group delay of $V =
\langle\beta_2\rangle \Delta\Omega_\mathrm{CS} L + \langle\beta_3\rangle \Delta\Omega_\mathrm{CS}^2 L/2\simeq 25$~fs
per round-trip for the CS. The situation is not as clear for the spectrum of Fig.~\ref{fig:spectra}(b), but we
estimate $V$ to be less than half that value in this case. With these values, the dispersion and loss coefficients
quoted above, and the detuning and driving power levels stated in the caption of Fig.~\ref{fig:spectra} (which then
determine the intracavity cw background power level $P_0$ based on the cavity mean-field model),
Eq.~(\ref{eq:dw_simplified}) predicts the emission of DWs at wavelengths (derived from the real part of~$Q$) of
$1540.9$--$1541.2$~nm (considering $V$ within 0--$12.5$~fs) for Fig.~\ref{fig:spectra}(b), and $1541.3$~nm for
Fig.~\ref{fig:spectra}(c). These predictions are in very good agreement with experiments and further support our
interpretation of the observed spectral peaks as genuine DWs.

\begin{figure}[t]
  \centerline{\includegraphics{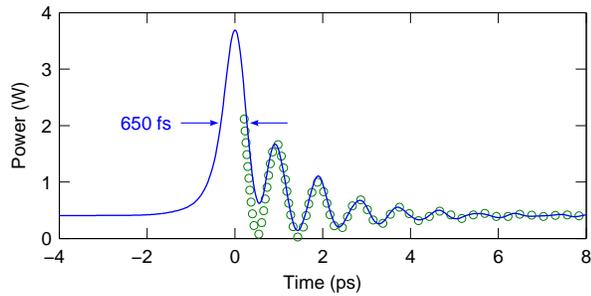}}
  \caption{Simulated intensity profile (blue) of the temporal CS corresponding to the spectrum shown in
    light-blue in Fig.~\ref{fig:spectra}(c), superimposed with the DW tail calculated with the analytical
    asymptotic analysis (green circles).}
  \label{fig:CSDW_temporal}
\end{figure}

Because the cavity losses are included in our analysis, the solutions of Eq.~(\ref{eq:dw_simplified}) are complex. By
examining the ansatz, it must be clear that the imaginary part of $Q$ is related to the time-constant with which the
dispersive radiation tail of the temporal CS exponentially decays to the background. We have unfortunately been
unable to temporally resolve this radiation in our experiment. Still, in order to provide some form of comparison, we
have plotted in Fig.~\ref{fig:CSDW_temporal} the intensity profile of the temporal CS (blue) corresponding to the
simulated (light-blue curve) spectrum plotted in Fig.~\ref{fig:spectra}(c), superimposed with the analytically
predicted DW tail (green circles). The latter corresponds to the expression $|E_0 + a \exp(-i Q \tau)|^2$ (ignoring
the weak down-frequency shifted component $b$ of the ansatz), with $Q=2\pi(1.07\ \mathrm{THz}) - i/(1.54\
\mathrm{ps})$ the solution of Eq.~(\ref{eq:dw_simplified}), and we used the complex DW amplitude $a$ as a fit
parameter. Clearly the agreement is excellent. Compounded by the good agreement between the simulated and
experimental data in Fig.~\ref{fig:spectra}, this strongly supports that we have observed  a temporal CS with a main
peak of about 650~fs duration, and with an oscillating DW tail decaying (in intensity) over a timescale of
about~3~ps.

In conclusion, we have directly observed experimentally the emission of DWs by temporal CSs. Our experiment was
performed in a dispersion-managed single-mode fiber ring resonator and the DWs manifest themselves as spectral peaks
that develop across the ZDW when the average cavity dispersion is sufficiently reduced. The peak becomes more
prominent and shifts closer to the driving beam wavelength with decreasing dispersion, in agreement with theoretical
analysis \cite{milian_soliton_2014}. Our observations further reinforce the case that Kerr combs generated in
microresonators are in some conditions underlined by temporal CSs, and could explain the very similar spectral peaks
observed in this context \cite{okawachi_octave-spanning_2011, coen_modeling_2013, erkintalo_coherence_2014}.

This work was supported by the Marsden Fund Council from Government funding, administered by the Royal Society of New
Zealand.

\clearpage

\section*{References with titles}

\end{document}